\documentstyle[floats,prl,aps]{revtex}
\input epsf
\begin{document}

\newcommand{\vertsp}{\vphantom{\displaystyle{\dot a \over a}}}
\newcommand{\se}{{(0)}}
\newcommand{\ve}{{(1)}}
\newcommand{\te}{{(2)}}
\newcommand{\nnu}{\nu}
\newcommand{\Spy}[3]{\, {}_{#1}^{\vphantom{#3}} Y_{#2}^{#3}}
\newcommand{\Gm}[3]{\, {}_{#1}^{\vphantom{#3}} G_{#2}^{#3}}
\newcommand{\Spin}[4]{\, {}_{#2}^{\vphantom{#4}} {#1}_{#3}^{#4}}
\newcommand{\scpot}{{\cal V}}
\newcommand{\tl}{\tilde}
\newcommand{\bm}{\boldmath}
\def\bi#1{\hbox{\boldmath{$#1$}}}

% redefine ell -> l for PRD
\renewcommand{\ell}{l}  
\renewcommand{\topfraction}{1.0}
\renewcommand{\bottomfraction}{1.0}
\renewcommand{\textfraction}{0.00}
\renewcommand{\dbltopfraction}{1.0}

%\int d\Omega\ (\Spin{Y}{s}{\ell}{m})(\Spin{Y}{s}{\ell'}{m'}{}^*)

\title{GRAVITATIONAL LENSING EFFECT ON COSMIC MICROWAVE BACKGROUND 
POLARIZATION}

\author{Matias Zaldarriaga${}^1$,
 	Uro\v s Seljak${}^2$}

\address{${}^1$ Department of Physics, MIT, Cambridge, MA 02139 \\ 
	${}^2$ Harvard Smithsonian Center For Astrophysics, Cambridge, 
	MA 02138}

\maketitle

\begin{abstract}
We investigate the effect of gravitational lensing by matter distribution
in the universe on the cosmic microwave background (CMB)
polarization power spectra and
temperature-polarization cross-correlation spectrum. 
As in the case of temperature spectrum
gravitational lensing leads to smoothing of narrow features 
and enhancement of power on the damping tail
of the power spectrum.
Because acoustic peaks in 
polarization spectra are narrower than in the temperature 
spectrum the smoothing effect is significantly larger and can 
reach up to 10\% for $l<1000$ and even more above that.
A qualitatively new feature is the generation of $B$ type polarization 
even when only $E$ is intrinsically present, such 
as in the case of pure scalar perturbations. This may be directly 
observed with Planck and other future small scale polarization experiments. 
The gravitational lensing
effect is incorporated in the new version (2.4) of CMBFAST code.
\end{abstract}

\section{Introduction}

Over the next few years a number of ground based, balloon and 
satellite experiments will measure CMB sky with an unprecedented
accuracy and detail.
The promise of a one percent precision
on the measured power spectrum of CMB anisotropies 
requires a similar accuracy in the theoretical predictions, if we
are to exploit all the information present in the data. The rewards
will be rich: among 
other things this will allow an accurate determination 
of a number of cosmological parameters and testing of current
structure formation theories
\cite{parameters}. In principle such a program is possible, since 
the anisotropies were produced when the universe was
still in the linear regime, which makes the calculations of model
predictions very accurate. In practice there are a number of 
important effects that need to be included
if this goal is to be realized. One of the most important among these
is the gravitational lensing effect.

As photons propagate through the universe from their last scattering 
to our detectors they are randomly deflected by the 
gravitational force exerted upon them 
by the inhomogeneous mass distribution. 
Previous work has shown
that gravitational lensing has an  effect on the
temperature anisotropy power spectrum which is 
not insignificant \cite{others,uroslens}. The
random deflections smear out the sharp features in the 
correlation function or power spectrum, leading to a suppression
of acoustic oscillations. Gravitational lensing can also enhance
power on the damping tail, causing it to decay less rapidly than 
predicted on very small angular scales
\cite{bentonsilk}. Gravitational lensing effect on the 
temperature anisotropies has been discussed several times in 
the literature and the formalism to calculate it using the 
evolution of density power spectrum both in linear and nonlinear regime 
has been presented in \cite{uroslens}.
In this paper we extend this calculation 
to the two linear polarization power
spectra and to the cross-correlation spectrum between temperature and
polarization. Because acoustic oscillations are narrower for polarization
spectra than for temperature, one expects gravitational lensing effect
to be more significant in the former and indeed our 
results confirm this. In addition, a 
qualitatively new effect is the mixing between $E$ and $B$ types of
polarization, which changes the pattern of polarization.
The outline of the paper is the following. 
In \S \ref{formalism} we develop
the formalism: this section contains
all the main analytic expressions needed 
for a numerical implementation of the effect. 
These have been numerically implemented in the new version of 
CMBFAST package (version 2.4) 
and require only a marginal increase in the CPU time
for their evaluations.
In \S \ref{estimate} we compute the effect for a typical
cosmological model and address the question of 
direct observability of the effect. We present the conclusions 
in \S \ref{conclusions}.   

\section{Two-point correlators in the presence of
lensing} \label{formalism}

The large scale density fluctuations in the universe 
induce random deflections in the direction of the CMB photons as they
propagate from the last scattering surface to us. This alters
the power spectrum  of both the temperature and polarization anisotropies.
The quantity is responsible for the deflections is the 
projected surface density.
Since the structures are not very correlated on large scales
the gravitational lensing effect 
is only relevant at the small angular 
scales in the CMB. We may hence use
the small scale limit formalism
\cite{uroslens}, which simplifies the calculations. 

The observed CMB temperature in the direction $\bi \theta$ is 
$T({\bi \theta})$ and
equals the unobservable temperature at the last scattering surface $
\tl T({\bi \theta}+\delta {\bi \theta})$, where
$\delta {\bi \theta}$ is the angular excursion of the photon as it 
propagates from the last scattering
surface until the present. 
In terms of Fourier components we have
\begin{eqnarray}
T({\bi \theta})&=&\tl T(\bi{\theta}+\delta{\bi \theta}) \nonumber \\
	 &=&(2\pi)^{-2}\int d^2{\bi l}\ 
e^{i{\bi l}\cdot (\bi{\theta}+ \delta
{\bi \theta})}\ T({\bi l}). 
\end{eqnarray}
The same relation applies to the two
Stokes parameters $Q$ and $U$ that describe linear polarization 
\cite{diegogab}.
In the small scale limit we 
take the direction of observation to
be near $\hat{\bi z}$ and we orient the local coordinate system 
orthogonal to $\hat{\bi z}$
to define $Q$ and $U$: $Q$ is the difference between
the photon intensities along $\hat {\bi x}$ and $\hat {\bi y}$, 
while $U$ is the difference between
photon intensities along the two diagonals.
In terms of Fourier components these can be expressed with 
opposite parity Fourier components $E({\bi l} )$ and 
$B({\bi l} )$ \cite{urospol}
\begin{eqnarray}
Q({\bi \theta})&=&\tl Q({\bi \theta}+\delta{\bi \theta}) \nonumber \\
	 &=&(2\pi)^{-2}\int d^2{\bi l}\  
e^{i{\bi l}\cdot ({ \bi \theta}+\delta { \bi \theta})}\
[E({\bi l} ) \cos(2\phi_{\bi l})
- B({\bi l}) \sin(2\phi_{\bi l})] \nonumber \\
U({\bi \theta})&=&\tl U({\bi \theta}+\delta {\bi \theta}) \nonumber \\
	 &=&(2\pi)^{-2}\int d^2{\bi l}\  
e^{i{\bi l}\cdot ({\bi \theta}+\delta {\bi \theta})}\
[E({\bi l}) \sin(2\phi_{\bi l})
+ B({\bi l}) \cos(2\phi_{\bi l})].
\label{expansion} 
\end{eqnarray}
The Fourier components satisfy
\begin{equation}
\langle \tl X({\bi l})\tl X({\bi l}^{\prime})\rangle = (2 \pi)^2\ C_{\tl Xl} \
\delta^{D}({\bi l}-{\bi l}^{\prime}),
\label{variance}
\end{equation}
with $\tl X=\tl T,\tl E,\tl B$ and the average is over different 
realizations of the CMB field.

The correlation function of the temperature between two points 
in the sky, ${\bi \theta^A}$ and ${\bi \theta^B}$,
only depends on their
angular separation ${\theta}$. 
With our choice of coordinate system the correlations between the 
polarization variables are also
a function of azimuthal angle $\phi$ \cite{urospol}. 
This is not the natural coordinate
system in which to define these correlation functions. Instead
one should align the local coordinate system
with the great circle
that connects the two points \cite{kamio}. 
With this choice parity conservation requires 
$T$ and $Q$ to be
uncorrelated with $U$ and the correlation functions
only depend on separation ${\theta}$. To obtain this set of correlation
functions in our coordinate system we can  calculate the
correlation between the variables at the origin 
and another point separated by an angle ${\theta}$ along
the ${\bi x}$ axis.
The correlation functions are calculated from equations
(\ref{expansion}) and (\ref{variance}),
\begin{eqnarray}
C_T({\theta})&=&\int {d^2{\bi l}\over (2\pi)^2}\   
e^{il\theta \cos \phi_{l}}
\langle e^{i{\bi l}\cdot (\delta{\bi \theta^A} -\delta{\bi
\theta}^{B})}\ \rangle \ \ C_{\tl Tl} \nonumber \\
C_Q({\theta})&=&\int {d^2{\bi l}\over (2\pi)^2}\  
e^{il\theta \cos \phi_{l}}
\langle e^{i{\bi l}\cdot (\delta{\bi \theta^A} -\delta{\bi
\theta}^{B})}\ \rangle\ \ [C_{\tl El} \cos^2(2\phi_{\bi l})
+C_{\tl Bl} \sin^2(2\phi_{\bi l})]\nonumber \\
C_U({\theta})&=&\int {d^2{\bi l}\over (2\pi)^2}\  
e^{il\theta \cos \phi_{l}}
\langle e^{i{\bi l}\cdot (\delta{\bi \theta^A} -\delta{\bi
\theta}^{B})}\ \rangle\ \ [C_{\tl El} \sin^2(2\phi_{\bi l})
+C_{\tl Bl} \cos^2(2\phi_{\bi l})] \nonumber \\
C_C({\theta})&=&\int {d^2{\bi l}\over (2\pi)^2}\
  e^{il\theta \cos \phi_{l}}
\langle e^{i{\bi l}\cdot (\delta{\bi \theta^A} -\delta{\bi
\theta}^{B})}\ \rangle\ \ C_{\tl Cl} \cos(2\phi_{\bi l}).
\label{corrlens}
\end{eqnarray}
The remaining average in equation
(\ref{corrlens}) is over the lensing fluctuations.
Only the cross-correlation between $Q$ and $T$ is
different from zero and we denote it $C_C({\theta})$. Even in the
presence of lensing $U$ does not cross correlate with either $T$ or
$Q$ because these quantities have opposite parities. 

The correlation function of the excursion angle can be used to
calculate the expectation value in equation (\ref{corrlens})
\cite{uroslens},
\begin{eqnarray}
\langle \exp\{ i{\bi l}(\delta{\bi \theta} -\delta{\bi
\theta}^{\prime})\}\ \rangle &=& \exp \{-{l^2 \over 2} [\sigma_0^2(\theta)
+ \cos(2 \phi_{\bi l}) \sigma_2^2(\theta)]\} \nonumber \\
&\approx& 1-{l^2 \over 2} [\sigma_0^2(\theta)
+ \cos(2 \phi_{\bi l}) \sigma_2^2(\theta)].
\label{avgexp}
\end{eqnarray}
The exponential above has been expanded out assuming $l^2 \sigma_2(\theta)$
and $l^2 \sigma_0^2(\theta)$ are small and we numerically 
verified this to be an excellent approximation (see also \cite{bernardau1}).
The two functions characterizing the rms dispersion of the photons
are \cite{uroslens}
\begin{eqnarray}
\sigma_0^2(\theta)&\equiv& 16 \pi^2 \int_0^\infty k^3 dk
\int_0^{\chi_{rec}} P_{\phi}(k,\tau=\tau_0-\chi)\ W^2(\chi,\chi_{rec})\
[1-J_0(k\theta\sin_K \chi)] \nonumber \\
\sigma_2^2(\theta)&\equiv& 16 \pi^2 \int_0^\infty k^3 dk
\int_0^{\chi_{rec}} P_{\phi}(k,\tau=\tau_0-\chi)\ W^2(\chi,\chi_{rec})\
J_2(k\theta\sin_K \chi). 
\label{sigmalens}
\end{eqnarray}
We denote with $\chi=\tau_0-\tau$ the comoving radial distance, $\tau$ is
the conformal time and $\tau_0$ corresponds to its value today. The 
comoving angular diameter distance is $\sin_K \chi=K^{-1/2}\sin K^{1/2}\chi$ in a closed universe ($K>0$),
$\chi$ in a flat universe ($K=0$) and $
(-K)^{-1/2}\sinh (-K)^{1/2}\chi$ in an open universe ($K<0$). The
curvature $K$
can be expressed using the present density
curvature parameter $\Omega_K=1-\Omega_m-\Omega_\Lambda$ and the present
Hubble parameter $H_0$ as $K=-\Omega_KH_0^2$.
The
power spectrum of the gravitational potential at time $\tau$ is
$P(k,\tau)$ and
$W(\chi,\chi_{rec})=\sin_K(\chi_{rec}-\chi)/\sin_K(\chi_{rec})$, 
where $\chi_{rec}$ is the radial distance to the last scattering surface 
at recombination.
We include the non linear evolution of the power spectrum using the
fitting formulae of Peacocks and Dodds \cite{peacock}. These expressions
are thus valid in a general Robertson-Walker metric, both in linear
and nonlinear regime. The nonlinear effects are however not very 
important except on small angular scales.

We use equations (\ref{avgexp}) and (\ref{sigmalens}) together with 
(\ref{corrlens}) to obtain the final expression for the correlation
functions,
\begin{eqnarray}
C_T(\theta)&=&\int {l dl\over 2\pi}
\ C_{\tl Tl}\ \{J_0(l\theta)[1-{l^2 \over 2}\sigma_0^2(\theta) ]
+{l^2 \over 2}\sigma_2^2(\theta)
J_2(l\theta)\}\  \nonumber \\ 
C_{Q}(\theta)+C_{U}(\theta)&=&\int {l dl\over 2\pi}
\ (C_{\tl El}+C_{\tl Bl})\ 
\{ J_0(l\theta)[1-{l^2 \over 2}\sigma_0^2(\theta) ]
+{l^2 \over 2} \sigma_2^2(\theta)
J_2(l\theta)\}   \nonumber \\ 
C_{Q}(\theta)-C_{U}(\theta)&=&\int {l dl\over 2\pi}
\ (C_{\tl El}-C_{\tl Bl})\ 
\{ J_4(l\theta)[1-{l^2 \over 2}\sigma_0^2(\theta) ]
+{l^2 \over 4} \sigma_2^2(\theta)
[J_2(l\theta)+J_6(l\theta)]\} 
 \nonumber \\ 
C_C(\theta)&=&\int {l dl\over 2\pi}
\ C_{\tl Cl}\ \{ J_2(l\theta)[1-{l^2 \over 2}\sigma_0^2(\theta) ]
+{l^2 \over 4} \sigma_2^2(\theta)
[J_0(l\theta)+J_4(l\theta)]\}  
. 
\label{fullcorr}
\end{eqnarray}

The power spectrum in Fourier space has become the most
widely used way to characterize the CMB anisotropies. We can compute the
different power spectra from the correlation functions,
\begin{eqnarray}
C_{Tl}&=&2\pi \int_0^\pi \theta d\theta \ C_T(\theta) \ J_0(l\theta) \nonumber \\
C_{El}&=&2\pi \int_0^\pi \theta d\theta \
\{ [C_Q(\theta)+C_U(\theta)]\ J_0(l\theta) 
+ [C_Q(\theta)-C_U(\theta)]\ J_4(l\theta) \} \nonumber \\
C_{Bl}&=&2\pi \int_0^\pi \theta d\theta \
\{ [C_Q(\theta)+C_U(\theta)]\ J_0(l\theta) 
- [C_Q(\theta)-C_U(\theta)]\ J_4(l\theta) \} \nonumber \\
C_{Cl}&=&2\pi \int_0^\pi \theta d\theta \
C_C(\theta)\ J_2(l\theta). 
\label{defcl}
\end{eqnarray}
Equations (\ref{fullcorr}) and (\ref{defcl}) give the mapping between
the observed CMB power spectra and the primordial one. Explicitly,
\begin{eqnarray}
C_{Tl}&=&C_{\tl Tl}+{\cal W}_{1 l}^{l^\prime}\ C_{\tl Tl^{\prime}}
\nonumber \\
C_{El}&=&C_{\tl El}+{1 \over 2}[{\cal W}_{1 l}^{l^\prime}+{\cal W}_{2
l}^{l^\prime}]\  C_{\tl El^{\prime}} + {1 \over 2}[{\cal W}_{1
l}^{l^\prime}-{\cal W}_{2 l}^{l^\prime}]\  C_{\tl Bl^{\prime}} 
\nonumber \\
C_{Bl}&=&C_{\tl Bl}+{1 \over 2}[{\cal W}_{1 l}^{l^\prime}-{\cal W}_{2
l}^{l^\prime}]\  C_{\tl El^{\prime}} + {1 \over 2}[{\cal W}_{1
l}^{l^\prime}+{\cal W}_{2 l}^{l^\prime}]\  C_{\tl Bl^{\prime}} 
\nonumber \\
C_{Cl}&=&C_{\tl Cl}+{\cal W}_{3 l}^{l^\prime}\ C_{\tl Cl^{\prime}},
\label{cll}
\end{eqnarray}
the sum over $l^{\prime}$ is implicit.
The window functions are defined to be
\begin{eqnarray}
{\cal W}_{1 l}^{l^\prime}&=&{{l^\prime}^3 \over 2}\int_0^{\pi} \theta d\theta
\ J_0(l\theta)\
\{\sigma_2^2(\theta)J_2(l^{\prime}\theta)
-\sigma_0^2(\theta)J_0(l^{\prime}\theta)\}
\nonumber \\
{\cal W}_{2 l}^{l^\prime}&=&{{l^\prime}^3 \over 2}\int_0^{\pi} \theta d\theta
\ J_4(l\theta)\
\{{1 \over 2}\sigma_2^2(\theta)
[J_2(l^{\prime}\theta)+J_6(l^{\prime}\theta)]
-\sigma_0^2(\theta)J_4(l^{\prime}\theta)\}
\nonumber \\
{\cal W}_{3 l}^{l^\prime}&=&{{l^\prime}^3 \over 2}\int_0^{\pi} \theta d\theta
\ J_2(l\theta)\
\{{1 \over 2}\sigma_2^2(\theta)[J_0(l^{\prime}\theta)
+J_4(l^{\prime}\theta)]-\sigma_0^2(\theta)J_2(l^{\prime}\theta)\}.
\label{clmapping}
\end{eqnarray}

Equations (\ref{cll}) and (\ref{clmapping}) are the main result of the paper. The
results for polarization are new and represent a generalization of
previous results for the temperature \cite{uroslens}. The important
qualitatively new feature is that lensing mixes
$E$ and $B$ polarization
modes. On small scales where the lensing effect is important all
cosmological models proposed so far predict only $E$ type
polarization and $C_{\tl B l}=0$. Lensing will however generate $B$
type polarization in the observed field, $C_{B l}={1 \over 2}[{\cal W}_{1
l}^{l^\prime}-{ \cal W}_{2 l}^{l^\prime}]\  C_{\tl El^{\prime}}\neq 0$.   
In the next section we will calculate the lensed power spectra in
a typical cosmological model to address the significance of the 
effect. Before addressing this issue let us explore
a simplified model to understand why the two
polarization types are mixed through lensing. 

So far we have introduced $E$ and $B$ type polarizations in Fourier
space. We can also define real space quantities
\begin{eqnarray}
E(\bi{\theta})&=&(2\pi)^{-2}\int d^2{\bi l}\ 
e^{i{\bi l}{\bi \theta}}\ E({\bi l}) \nonumber \\
B(\bi{\theta})&=&(2\pi)^{-2}\int d^2{\bi l}\ 
e^{i{\bi l}{\bi \theta}}\ B({\bi l}).
\end{eqnarray}
These two quantities describe completely the polarization field
and it proves easier to understand the effect of lensing in terms of these.
They can also be computed directly from $Q$ and $U$ in
real space \cite{theses},
\begin{eqnarray}
E(\bi{\theta})&=&\int d^2{\bi \theta}^{\prime}\ 
\omega(|\bi{\theta}^{\prime}-\bi{\theta}|)\ Q_r(\theta^{\prime}) \nonumber \\
B(\bi{\theta})&=&\int d^2{\bi \theta}^{\prime}\ 
\omega(|\bi{\theta}^{\prime}-\bi{\theta}|)\ U_r(\theta^{\prime}).
\label{ebreal}
\end{eqnarray}
We have defined $Q_r$ and $U_r$, the Stokes parameters in the
polar coordinate system centered at $\bi{\theta}$. 
If $\bi{\theta}=0$ then $Q_r=\cos 2\phi^{\prime}\
Q(\bi{\theta}^{\prime}) - \sin 2\phi^{\prime}\
U(\bi{\theta}^{\prime})$ and $U_r=\cos 2\phi^{\prime}\
U(\bi{\theta}^{\prime}) + \sin 2\phi^{\prime}\
Q(\bi{\theta}^{\prime})$. The window is $w(\theta)=1/\theta^2\;
(\theta\neq 0)$, $w(\theta)=0 \;
(\theta= 0)$. 

We consider as a toy model an unlensed  polarization field which is a radial
pattern around the origin in a ring of size $\theta_0$. We have
$\tilde Q(\bi{\theta})=E_0  \theta_0
\delta^D(\theta-\theta_0)/2\ \cos 2\phi$ and 
$\tilde U(\bi{\theta})=E_0  \theta_0
\delta^D(\theta-\theta_0)/2\ \sin 2\phi$. In the absence of lensing we
would observe at the origin $\tilde E(\bi{\theta}=0)=E_0$ and $\tilde
B(\bi{\theta}=0)=0$. This follows from $U_r=0$ (because 
polarization is radial) and 
equation (\ref{ebreal}). We now
consider what happens if we shift the position of the photons by a
random angle of size $\delta \theta$. Each segment of the ring will be
mapped to a different position with angle $\phi$ where
$|\delta\phi|\sim |\delta \theta|/\theta_0$. Each segment acquires 
a random
component of non radial polarization and since we are assuming 
the segment shifts are uncorrelated
the integral of $U_r$ over $\phi$ does not vanish. It has
mean zero and
variance $\langle U_r^2\rangle \propto (\delta \theta/\theta_0)^2 \
Q_r^2$. The  
measured power in the  $B$ mode is $\langle B^2(\bi{\theta}=0)
\rangle\propto(\delta \theta/\theta_0)^2 E_0^2$. This example shows 
that the 
pattern of polarization vectors on the sky determines the
amount of $E$ and $B$ type polarization. 
Lensing distorts this pattern by shifting the positions of the photons
in the plane of the sky relative to the last scattering surface. It  
can thus generate $B$ pattern out of initially pure $E$ polarization. 
In the next section we discuss the amplitude of this effect.

\section{Estimate of the lensing effect} \label{estimate}

\begin{figure*}
\begin{center}
\leavevmode
\epsfxsize=6in \epsfbox{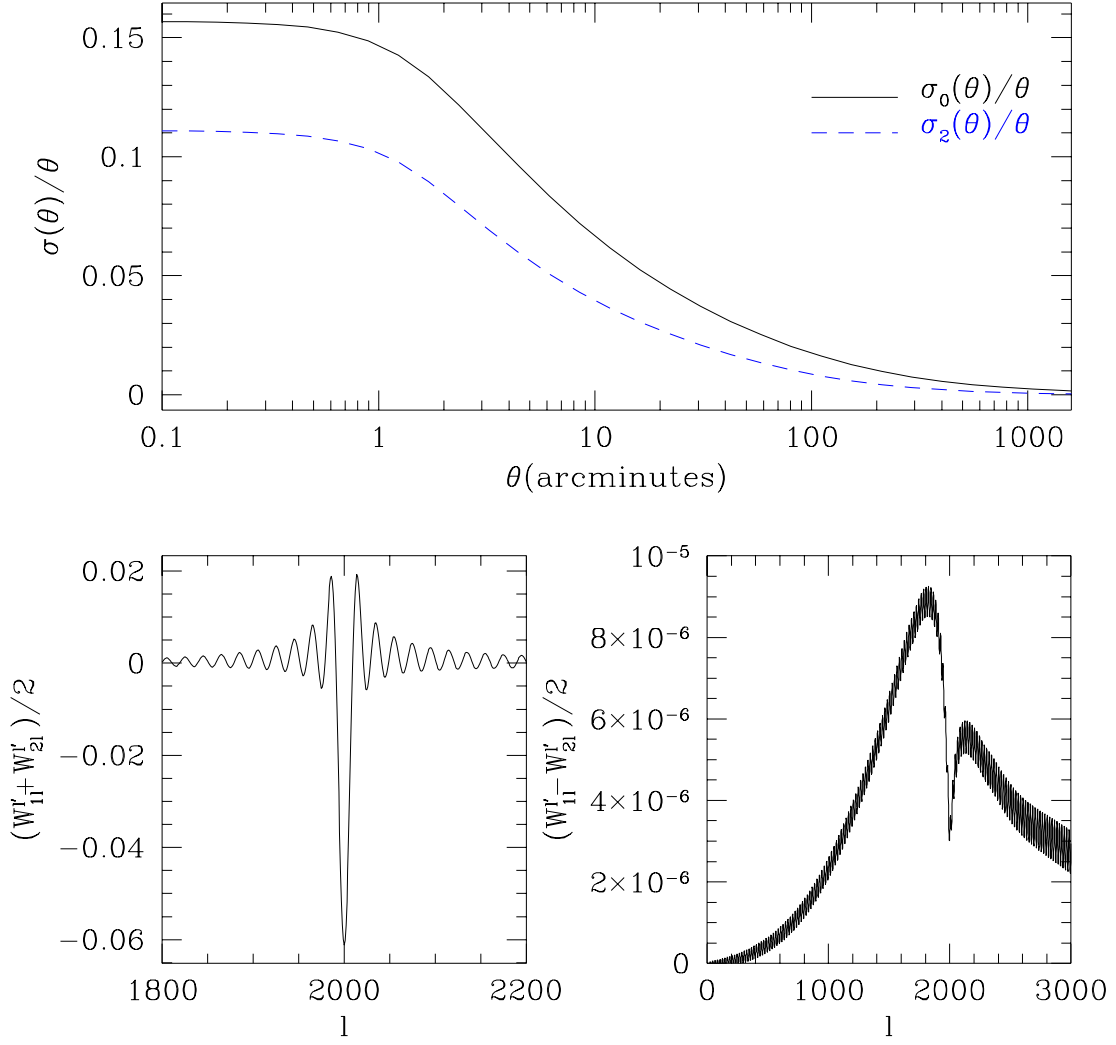}
\end{center}
\caption{The upper panel shows the two functions $\sigma_0(\theta)$
and $\sigma_2(\theta)$ for the cosmic concordance 
model discussed in the text. The lower
panels show ${1 \over 2}[W_{1 l}^{l^{\prime}} \pm W_{2 l}^{l^{\prime}}]$ 
for $l=2000$. The Window function for the mixing between $E$ and $B$ is
is not well localized in $l$ space and is rapidly oscillating as a 
function of $l$. } 
\label{fig1}
\end{figure*}

The effect of gravitational lensing on the temperature power spectrum has been
studied in detail in previous work \cite{uroslens}. 
We will explore here
the effects on polarization and asses the detectability of the $B$
signal. In these examples we will use the cosmic concordance 
model \cite{ostein} with $H_0=65\ $ km s$^{-1}$Mpc$^{-1}$,
$\Omega_\Lambda=0.65$, $\Omega_m=0.35$, $\Omega_b h^2=0.015$ and
$n=1$. The model is COBE normalized and satisfies the constraints
on 8 h$^{-1}$Mpc scale from the abundance of clusters. 

Figure \ref{fig1} shows the functions that characterize the rms
deflection of the photons, $\sigma_0(\theta)$ and
$\sigma_2(\theta)$. On small scales the rms relative deflection between 
two photons approaches 15\%, justifying the weak lensing assumption over
most of the sky (for caustic formation one requires $\delta \theta \sim 
\theta$, which will occur only on rare occasions).
We also show  ${ 1 \over 2}[{\cal W}_{1 l}^{l^\prime} \pm 
{\cal W}_{2 l}^{l^\prime}]$ for
$l=2000$.
The two
window functions  ${\cal W}_{1 l}^{l^\prime}$ and ${\cal
W}_{2 l}^{l^\prime}$ are very similar, making their difference
much smaller than their sum. One thus expects the
generated $B$ type polarization from a pure $E$ type (and viceversa) 
to be very small. In fact ${\cal W}_{3 l}^{l^\prime}$ is
also very similar, the three windows only differ at the 1\% level.
The windows are oscillatory 
as a function of  $l^{\prime}$ and
their main contribution is concentrated around $l$. 

\begin{figure*}
\begin{center}
\leavevmode
\epsfxsize=6in \epsfbox{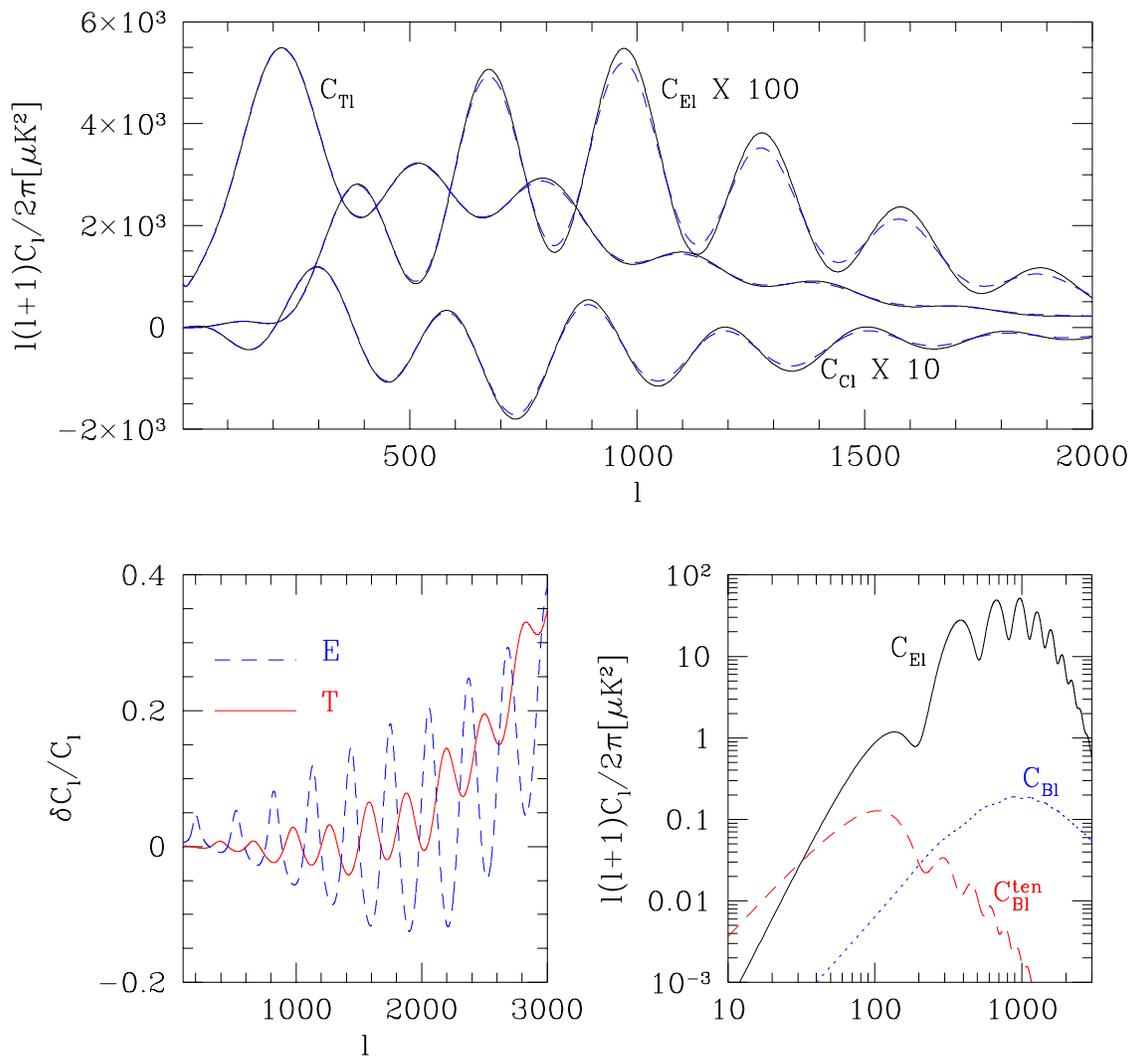}
\end{center}
\caption{The upper panel shows the $T$, $E$ and $C$ power spectra. Dashed (solid)
lines correspond to the lensed (unlensed) spectra. The bottom left
panel shows the relative difference between the lensed and unlensed
spectra for $T$ and $E$ ($\delta
C_l/C_l \equiv (C_l^{lensed}-C_l^{unlensed})/C_l^{unlensed}$) and shows 
both suppression of oscillations and enhancement of power on small scales. 
The bottom right panel shows the $B$ type
polarization induced by lensing. 
We include
the $E$ and $B$ spectra for inflationary 
model where scalars and tensors produce
equal amount of power in the temperature on COBE scales ($T/S=1$).}
\label{fig2}
\end{figure*}

Figure \ref{fig2} (upper panel) 
shows the lensed and unlensed power spectra. 
On intermediate scales lensing has the general effect of smearing
the peaks in the spectrum by redistributing power. Polarization 
receives contribution only from velocity gradients at the last scattering
surface and the acoustic peaks are very sharp. In contrast, temperature
receives contributions both from velocity and density of photon-baryon 
plasma and the two are out of phase with each other, leading to 
a partial cancellation of acoustic peaks. 
One thus expects the effect of lensing to be larger for polarization and
indeed we find that it is
approximately a factor of two larger for $E$ polarization 
(figure \ref{fig2}, bottom left panel). 
%We ignore the gravitational lensing effect
%on $B$ type polarization, since it can only be generated on large 
%angular scales (where tensor modes contribute to it) and is so 
%negligibly affected by gravitational lensing.
In the damping tail the power is
enhanced over the unlensed case as can be seen in the general
trend in the relative difference between lensed and unlensed spectra 
(figure \ref{fig2}). 

To asses the detectability of the induced $B$ polarization we will
focus on the Plack mission. We 
assume that measurements of the temperature, $E$ type polarization and
$T-E$ cross correlation 
have allowed the  determination of the cosmological
parameters accurately enough so that we know the approximate shape of
the $B$ power spectra induced by lensing. We will attempt to determine just one
parameter, the overall normalization of this $B$ signal, which expresses
the overall effect of the matter fluctuations along the line of sight. 
If the signal to noise were large enough we could also attempt to 
determine the power spectrum of the lensing effect, by exploring the 
signal in $B$ as a function of $l$.

The relative error on the overall amplitude of the induced $B$ component $\beta$ is
\begin{equation}
{\Delta\beta \over \beta}=\sqrt{2\over f_{sky}\sum_l
(2l+1)/(1+w_P^{-1} B_l^{-2}/C_{Bl})^2};
\end{equation}
$f_{sky}$ is the fraction of observed sky which we take to be
$f_{sky}=0.8$. We assume the noise and beam width to be
$w_P^{-1}=(0.025\ \mu K)^2$ and $B_l^{-2}=e^{l^2 \sigma_b^2}$ with
$\sigma_b=\theta_{fwhm}/2 \sqrt{2 \ln 2}=9\times 10^{-4}$. The effective
number of modes contributing information is $N_{eff}=f_{sky}\sum_l
(2l+1)/(1+w_P^{-1} B_l^{-2}/C_{Bl})^2$, so that $\Delta
\beta/\beta=\sqrt{2/N_{eff}}$. 
For the experimental specifications above $N_{eff}\approx 8$ 
and $\Delta \beta/\beta \approx 0.5$. This means that there are 
only 8 independent modes of $B$ type polarization
that can be observed with Planck, each a gaussian random variable
with 0 mean and so the signal will be at the limit of 
detectability by Planck. The $B$ signal peaks at $l=1000$ and a ground
based experiment observing a small patch of the sky for sufficiently 
long time to reduce the
noise per pixel would be more effective to detect this signal. Gravitational
lensing induced $B$ is also not a significant contaminant of the $B$
polarization expected from tensor modes in inflationary models \cite{gravity}, the 
latter being dominant on large angular scales (figure \ref{fig2}, bottom
right panel, where we assumed tensors and scalars are of equal 
amplitude in the temperature on large scales).

\section{Conlusions} \label{conclusions}

The change in the CMB spectra induced by random deflections of the
photons by the large scale structure of the universe has to be
included in the calculations of the anisotropies when comparing theory
and observations.
We have presented a formalism to calculate the effect of gravitational 
lensing on
all four CMB power spectra. The final expressions are given in a 
compact and numerically efficient form that is adequate for numerical
implementation. These expressions have
been implemented in the CMBFAST package, freely available from the 
authors. The computational time does not significantly increase with
this feature and since the effect can be significant we feel the 
lensing effect should be included in any calculation where high 
precision accuracy is important, such as in the design and 
analysis of the CMB experiments. The method is self-consistent
in the sense that for any cosmological model we use the actual power 
spectrum computed in the code to compute the lensing effect. The 
power spectrum is normalized to COBE using the CMB spectrum computed
from the same code output. This approach gives the correct amplitude 
of gravitational lensing effect for the particular model in question.
It should be pointed out that 
if the model is inconsistent with the small scale constraints
such as $\sigma_8$ normalization then the amplitude of the lensing effect
will also be incorrect. For example, COBE normalized standard CDM has a 
much larger gravitational lensing effect than what we find using the
cosmic concordance model, but this is only a reflection of standard CDM
model having too much small scale power to be compatible with small 
scale constraints. For models that are correctly
normalized on small scales 
the relative change of the polarization power
spectra can reach 10\% at $l \sim 1000$ and even more at higher $l$.
This is larger than in the temperature spectrum and is caused by the
sharper acoustic peaks in the polarization spectra. In the damping 
region the lensed spectra show an enhancement above the unlensed spectra 
just like in the temperature case,
although the sensitivity to this effect is too small for satellite missions 
to detect it in polarization \cite{gravity}.

Gravitational lensing also mixes 
$E$ and $B$ type polarization by deforming the
polarization pattern on the sky relative to that at the last
scattering surface. This will generate $B$ type polarization out
of the $E$ type polarization even if there was no $B$ present at 
the last scattering surface. This induced $B$ mode is rather
small in typical models 
and will be only marginally 
detectable by the Planck Surveyor. It peaks at fairly small angular 
scales around
$l\sim 1000$ and so does not affect the measurement of gravity waves 
from $B$ polarization on larger scales. A ground based experiment observing
a small patch of
the sky would be more suitable to observe this effect and would 
allow one to determine the power spectrum of matter fluctuations.
This would allow a detection of a combined gravitational 
effect from structures 
spanning a much larger range in redshift space 
than currently reachable by other methods.
\section*{Acknowledgments}
MZ was supported by NASA grant NAG5-2816.

\end{document}